# Substate Profiling for Effective Test Suite Reduction

Chadi Trad, Rawad Abou Assi, Wes Masri
Electrical and Computer Engineering Dept.
American University of Beirut
rawad84@gmail.com, {wm13, cht02}@aub.edu.lb

*Abstract*—Test suite reduction (TSR) aims at removing redundant test cases from regression test suites. A typical TSR approach ensures that structural profile elements covered by the original test suite are also covered by the reduced test suite. It is plausible that *structural profiles* might be unable to segregate failing runs from passing runs, which diminishes the effectiveness of TSR in regard to defect detection. This motivated us to explore *state profiles*, which are based on the collective values of program variables.

This paper presents *Substate Profiling*, a new form of state profiling that enhances existing profile-based analysis techniques such as TSR and coverage-based fault localization. Compared to current approaches for capturing program states, *Substate Profiling* is more practical and finer grained.

We evaluated our approach using thirteen multi-fault subject programs comprising 53 defects. Our study involved greedy TSR using *Substate* profiles and four structural profiles, namely, basic-block, branch, def-use pair, and the combination of the three. For the majority of the subjects, *Substate Profiling* detected considerably more defects with a comparable level of reduction. Also, *Substate* profiles were found to be complementary to structural profiles in many cases, thus, combining both types is beneficial.

I. INTRODUCTION

*Test suite reduction* (*TSR*), also referred to as *test suite minimization* [31][45][56], aims at reducing the number of test cases of a given regression test suite in a manner that does not compromise its defect detection ability [45][46][52][53]. Naturally, there is cost incurred when performing TSR. However, in many situations, such cost would be insignificant compared to the cost savings yielded by the reduced test suite. In particular, TSR is most beneficial under one or both of the following circumstances: 1) *test suite execution is highly costly*, e.g., due to manual effort or resource consumption; and 2) *manual auditing is required*, i.e., the tester is required to manually determine whether test cases pass or fail. Rothermel *et al.* [47] reported of a case in industry where executing the complete regression test suite (of a 20K LOC product) required seven weeks. *Cyber Physical Systems* [39][40][41] is an example domain where TSR might be highly beneficial. In such domain, software closely interacts with physical processes; therefore, test execution is typically costly, and test oracles are likely to be absent, thus requiring manual auditing. On the other hand, TSR might not be worthwhile in situations where test oracles are fully automated and test suite execution is repeatable at no major additional cost.

A widely adopted approach for test suite reduction ensures that the test requirements [1] satisfied by the original test suite are also satisfied by the reduced test suite [45][51][58]; in other words, both test suites are required to cover the same program elements. Commonly used program elements include *methods*, *statements*, *branches*, and *def-use pairs* [21]; which are structural in nature and relatively simple [31][38][58]. More complex and seldom used structural program elements include acyclic paths [4], call-stacks [42], and slice pairs [38].

It is plausible that a given failure might not be characterized by the execution of any structural program element, be it simple or as complex as a *complete path* [1]. That is, there are situations in which *structural execution profiles*[1] are unable to segregate failing runs from passing runs, or from other failing runs that are due to different defects. In such situations, test suite reduction is likely to yield reduced test suites that are not as effective as their respective original test suites in regard to defect detection. This motivated us to explore alternative types of profiles, namely, state profiles that are constructed based on the collective values of program variables. Similar to structural profiles, state profiles are meant to capture the execution behavior of programs, but from a memory perspective. Thus, it is more appropriate to define a *state profile* as being data representing the memory behavior of an executing program.

Researchers have devised numerous test suite reduction techniques of which most could be classified as greedy [9][29][46], heuristic-based [20][7][8], distribution-based [29][38], or ILP-based [22][31]. This paper is not proposing a new test suite reduction technique, but a new type of profiling that makes existing profile-based reduction techniques more effective. Specifically, this paper presents *Substate Profiling*, a new form of state profiling that is relatively practical in addition to being fine grained with respect to what variables to consider and when/where to record their values. *Substate Profiling* is suitable for profile-based dynamic analysis techniques such as test suite reduction and prioritization, and coverage-based fault localization. To assess the benefits of using *Substate Profiling*, we conducted a comparative study involving greedy test suite reduction using *Substate* profiles and four other structural profiles, namely, basic-block, branch, def-use pair, and the combination of the three (termed *ALL* [38]). The experiments involved thirteen multi-fault Java programs, which included a total of 53 defects. The results showed that, comparatively, *Substate Profiling* detected considerably more defects for the majority of the subjects, while exhibiting comparable reduction

---

[1] An *execution profile* comprises recorded information meant to capture the runtime behavior of a program during a given test run

levels in most cases. Our results also showed that in many cases *Substate* profiles are complementary to structural profiles, thus, combining them is beneficial.

Work related to *Substate Profiling* entails capturing and comparing program states for various purposes. Zimmerman and Zeller [59] modeled memory as a graph that captures program states to assist debugging. Xie and Notkin [55] captured the values of global variables and function parameters in order to compare state profiles, which they termed "value spectra". Xie *et al.* [54] minimized unit test suites by capturing the state of objects at the entry of test functions. Elbaum *et al.* [15] analyzed program states in order to create tests that are hybrid of unit and system tests. Jaygarl *et al.* [25] proposed OCAT, an approach that captures, generates, and mutates objects in order to improve *Randoop*. Francis [16] collected samples of object states to enable greedy test suite reduction and prioritization techniques.

The aforementioned body of work differs from ours primarily in regard to the granularity at which program states are captured. *Substate Profiling* is the only one aiming at capturing/approximating the memory behavior throughout program execution. Whereas existing techniques record states at the start/end of tests or functions, we record state information at every definition statement in addition to start/end of functions. Section VI further details these differences and presents more related work. The main contributions of this work are as follows:

- A new state profiling approach, termed *Substate Profiling*, which is fine grained and suitable for profile-based dynamic analysis techniques such as test suite reduction and prioritization, and coverage-based fault localization
- Supporting tools for the Java platform
- An experimental study that contrasts the effectiveness of using *Substate* profiles to that of using structural profiles in greedy test suite reduction
- Insight of whether *Substate* profiles are complementary to structural profiles

The remainder of this paper is organized as follows. Section II describes commonly used structural profiles. Section III walks through a motivating example. Section IV presents and illustrates our proposed *Substate Profiling* approach and tools. Section V reports on our empirical study and summarizes our findings. Section VI discusses related work, and Section VII presents our conclusions and future work.

## II. BACKGROUND

This section presents the four structural profiles involved in our experiments presented in Section V. We use the same terminology adopted in [38] to describe them.

- Basic blocks (*BB*): For every basic block *B* such that *B* is executed in at least one test, a *BB* profile indicates (via a 0 or 1 entry) whether *B* is executed in the current test.
- Basic-block edges (*BBE*) or branches: For every pair of basic blocks *B1* and *B2* such that there is a branch from *B1* to *B2* in at least one test, a *BBE* profile indicates whether this branch is taken in the current test.
- Def-use pairs (*DUP*): For each pair of statements $s_1$ and $s_2$ such that: 1) $s_1$ defines a variable $x$; 2) $s_2$ uses $x$; and 3) $s_1$ dynamically reaches $s_2$ in at least one test; a *DUP* profile indicates whether $s_1$ dynamically reaches $s_2$ in the current test.
- All above profiles combined (*ALL*): Combined entries of *BB, BBE* and *DUP*.

In order to generate the above structural profiles, we built a tool that targets the Java platform based on the ASM Java bytecode manipulation and analysis framework (*asm.ow2.org*).

## III. MOTIVATING EXAMPLE

As noted earlier, there are situations in which structural profiles fall short at characterizing defects. This section provides an example[2] that demonstrates a case where state profiling performs better than structural profiling. Consider the Java program shown in Table 1. Given a string representing an eight-digit binary number, the function decimal() is meant to return its decimal conversion. When variable i is zero, statement 8 causes variable increment to overflow and take on the value -128 as opposed to 128 (since the range of byte is [-128, 127]); thus causing a failure whenever the input string has its leftmost bit set. Table 1 also shows six test cases, two of which trigger a failure, and their corresponding statement coverage information: a check mark indicates that the statement at the given row was executed at least once using the test case at the given column (i.e., the profile count was non-zero). Note how the resulting statement profiles for both passing and failing test cases are identical, and thus are not helpful in techniques such as test suite reduction or fault localization.

Table 2 and Table 3 respectively show the branch coverage and def-use coverage information. Here also, all test cases exhibit the same profiles, which deems them not useful. Therefore, *BB*, *BBE*, *DUP* (and *ALL*) structural profiles fail to differentiate between the passing and failing runs in our example.

We now shift our focus to state profiling by zeroing in on statement 6 where variable increment is repeatedly defined. Figure 1 shows the values taken by increment throughout the execution of the test cases. Clearly, the curves for the two failing test cases $t_5$ and $t_6$ exhibit shapes that are considerably different from the rest, suggesting that in our example, state profiling would be more useful than structural profiling. It is worth noting that profiling the values taken by variable decimal at statement 7 would also help differentiate the failing test cases from the passing ones.

The goal of this work is to devise a profiling approach and tool that capture anomalous state behaviors such as those exhibited by increment and decimal.

## IV. SUBSTATE PROFILING: APPROACH, TOOLS, AND APPLICATIONS

The purpose of a profiling tool is to detect and record some events of interest that occurred during a program execution. In the context of structural profiling, an event is the execution of a program element, e.g., a statement or branch. In the context of state profiling, an event is the instantiation of a program state. Which brings forward two critical questions:

1) What constitutes a *program state*, is it a value-snapshot of all program variables?
2) When should the program state be captured, should it be done following the execution of every definition statement?

---

[2] This example is borrowed from the authors' previous related work [34].

| /*Given a string representing an 8 digit binary number, the method decimal() returns its decimal conversion. Due to overflow, failure occurs whenever the input string has its leftmost position set */ <br> public class BinarytoDecimal { | Passing | | | | Failing | |
|---|---|---|---|---|---|---|
| | $t_1$ | $t_2$ | $t_3$ | $t_4$ | $t_5$ | $t_6$ |
| | 00101111 | 10111010 | 01111100 | 10111110 | 11101111 | 10101101 |
| public static void main(String args[]) { | | | | | | |
| 1  decimal(args[0]); } | ✓ | ✓ | ✓ | ✓ | ✓ | ✓ |
| public static int decimal(String binary) { | | | | | | |
| 2  int decimal = 0; | ✓ | ✓ | ✓ | ✓ | ✓ | ✓ |
| 3  for (int i = 0; i < binary.length(); i++) { | ✓ | ✓ | ✓ | ✓ | ✓ | ✓ |
| 4  byte increment = 0; <br> 5  if (binary.charAt(i) == '1') { | ✓ | ✓ | ✓ | ✓ | ✓ | ✓ |
| 6  increment = (byte)Math.pow(2.0,(double)(7-i)); | ✓ | ✓ | ✓ | ✓ | ✓ | ✓ |
| } <br> 7  decimal += increment; | ✓ | ✓ | ✓ | ✓ | ✓ | ✓ |
| } <br> 8  return decimal;}} | ✓ | ✓ | ✓ | ✓ | ✓ | ✓ |

**Table 1** – Java Code and BB coverage

| Branch Coverage | | | | | | |
|---|---|---|---|---|---|---|
| (src, target) | P | | | | F | |
| | $t_1$ | $t_2$ | $t_3$ | $t_4$ | $t_5$ | $t_6$ |
| (3,4) | ✓ | ✓ | ✓ | ✓ | ✓ | ✓ |
| (3,8) | ✓ | ✓ | ✓ | ✓ | | |
| (5,6) | ✓ | ✓ | ✓ | ✓ | ✓ | ✓ |
| (5,7) | ✓ | ✓ | ✓ | ✓ | ✓ | ✓ |

**Table 2** – Branch coverage

| Def-Use Coverage | | | | | | |
|---|---|---|---|---|---|---|
| (var, def, use) | P | | | | F | |
| | $t_1$ | $t_2$ | $t_3$ | $t_4$ | $t_5$ | $t_6$ |
| (i,3,5) | ✓ | ✓ | ✓ | ✓ | ✓ | ✓ |
| (binary,1,3) | ✓ | ✓ | ✓ | ✓ | ✓ | ✓ |
| (binary,1,5) | ✓ | ✓ | ✓ | ✓ | ✓ | ✓ |
| (increment,4,7) | ✓ | ✓ | ✓ | ✓ | ✓ | ✓ |
| (increment,6,7) | ✓ | ✓ | ✓ | ✓ | ✓ | ✓ |
| (decimal,2,7) | ✓ | ✓ | ✓ | ✓ | ✓ | ✓ |
| (decimal,7,8) | ✓ | ✓ | ✓ | ✓ | ✓ | ✓ |

**Table 3** – Def-use coverage

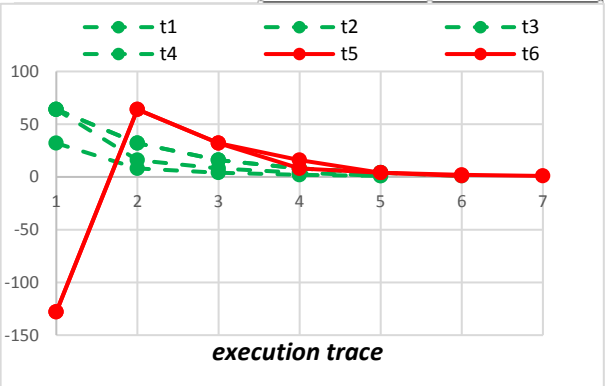

**Figure 1** – Values taken by increment

To achieve ultra-high accuracy, the answer should be *yes* for both questions. However, this is clearly not feasible, which calls for an approximating approach. In fact, as presented in Section VI, we are not aware of any related work that does not involve some form of an approximation strategy.

Our strategy for making *Substate Profiling* feasible must not compromise its ability to characterize non-trivial failures. For that purpose, our design guarantees that *Substate Profiling* captures the data of interest following every event that can potentially change the program state. Specifically, following the execution of every definition statement (in addition to other locations as discussed later). However, for the sake of scalability, the captured data does not comprise a value-snapshot of all program variables, but only the values of the variables being assigned at the capture location. In other words, instead of considering the overall-states exhibited by a program *P*, *Substate Profiling* considers the collective *sub-states* exhibited at capture locations in *P*, specifically, at definition statements, return statements, and at the entry of functions (in order to capture the values of formal parameters). Hereafter, we will refer to these locations as *capture points* or *Cp*'s for short [19].

Our proposed approach comprises three main components: the *DataCollector*, the *FeatureExtractor*, and the *ProfileGenerator*. Given a program *P* and an associated test suite *T*:

1) For each test case *t* in *T*, the *DataCollector* is applied in order to capture the data at the *Cp*'s in *P*.
2) For each test case *t* in *T*, the collected raw data at each *Cp* is abstracted into features using the *FeatureExtractor*. Therefore, if a *Cp* is executed by *n* different test cases, it will be associated with *n* sets of features.
3) At each *Cp*, the *ProfileGenerator* identifies one or more profile elements. This is achieved by applying cluster analysis based on the extracted features. Hence, a profile element at a *Cp* would represent a substate behavior exhibited by a subset of the *n* test cases that executed that *Cp*.
4) As a final step, the *ProfileGenerator* uses the complete set of profile elements (gathered from all the *Cp*'s) to build a *Substate* profile for each test case. This step is straightforward since the mapping between a profile element and the test cases that exhibited its behavior is readily available.

Next, we provide a detailed description of the *DataCollector*, the *FeatureExtractor*, and the *ProfileGenerator*. Then we provide an illustrative example and discuss potential applications of our profiling approach. (It might be helpful for some readers to first skim through the illustrative example presented in Section IV.D).

A.   *DataCollector*

The *DataCollector* is applied to each test case in order to capture the data at the *Cp*'s, i.e., definition statements, return statements, and functions entries. To better match a given *Cp* across test cases, we associate it with the following identifying attributes: 1) the signature of the method it belongs to; 2) the instruction offset within the method; and 3) the identifier of the thread within which it executed. Therefore, in order to compare the substates induced at a given *Cp* across a set of test cases, we require the three aforementioned attributes to be identical for all of the test cases in the set.

In case of definition statements, the variables being defined at a *Cp* could be local variables, formal parameters, static variables, attributes of class instances, or array elements. When the data at a *Cp* is of primitive scalar type we record its value as is, and in case it is of *java.lang.Number* type we record the value returned by *doubleValue()*. This scalar data can be used directly by the *FeatureExtractor*, described in the next section. Whereas, in case the data is of type *java.lang.String*, the captured strings need to be first mapped to numeric measures in order to extract meaningful features characterizing them. We opted to use three measures to represent a given string, namely, its *length*, *richness* [11][6], and *entropy* [14]. (We explored using the string's hash code, but realized that it is not appropriate since using it to compute most of the features listed in Section IV.B would yield values that do not characterize strings in any meaningful way.)

Special consideration was given to the data collected at a *Cp* if it included one or more *NaN* or *Infinity* values, as described in Section IV.C. Also, the current implementation of the *DataCollector* considers an array element definition to be a definition of the entire array; i.e., when recording data values, it does not take into consideration the array indexes. This design decision was made for efficiency purposes, however, we intend to revisit it in future work.

As detailed later, the *ProfileGenerator* does not operate on the raw data collected by the *DataCollector*, but on features extracted from that data. Therefore, the data collected at each *Cp* needs to be analyzed (for each test case) in order to extract features that compactly characterize it. Some features could be determined on the fly, and thus are computed within the *DataCollector* as the data points are being collected; we will refer to this set of *simple* features as $F_{simple}$. Other features require all the data points to be present a priori, for those, the *DataCollector* stores the data until the end of execution in order to compute them; we will refer to this set of features as $F_{complex}$.

Clearly, if we only consider the features in $F_{simple}$ then there would not be any concerns about memory consumption regarding the *DataCollector*. However, in order to consider the features in $F_{complex}$, we need to take precautionary measures since we collect every value that gets exhibited at every *Cp*. Specifically, at each *Cp*, we opted to store the first $V_{lead}$ leading, and the $V_{trail}$ trailing captured data/value points, where $V_{lead}$ and $V_{trail}$ are configurable (in our experiments we use $V_{lead} = V_{trail} = 2000$). It should be noted that we do not totally discard the (middle) data points that were not stored since we use them in computing the features in $F_{simple}$.

The current implementation of the *DataCollector* comprises 3,168 lines of Java code; it is composed of two components: the *Instrumenter* and the *Profiler*. The preliminary step in applying the *DataCollector* is to instrument the target byte code classes and/or jar files using the *Instrumenter*, which was also implemented using the ASM framework (*asm.ow2.org*). The *Instrumenter* inserts a number of method calls to the *Profiler* at the following points of interest:

- Methods entries: to capture the values of formal parameters.
- Methods exits: to capture the return values.
- Definition statements: to capture the values assigned to local variables, static variables, attributes of class instances, and array elements.
- Additional points of interest (e.g., throw statements) to enable better monitoring of the subject programs.

At runtime, the instrumented application invokes the *Profiler*, passing it information that enables it to collect values assigned at the *Cp*'s. It is worth noting that the *DataCollector* is thread-safe, which is particularly necessary for tracking local variables.

### B. FeatureExtractor

For each test case, fourteen features are extracted from the data collected at each *Cp* to be passed on to the *ProfileGenerator*. These features are meant to compactly abstract the state behavior at a given *Cp*. Seven of these features, which constitute $F_{simple}$, are computed on the fly by the *DataCollector*, namely, *Size*, *Min*, *Max*, *Mean*, *Decreasing*, *Increasing*, and *longestRunOfZeros*.

The remaining features, which constitute $F_{complex}$, are computed at the end of each profiled test execution. They include *Median*, *Mode*, *Standard deviation*, *Inter-quartile range*, *Skewness*, *Kurtosis* and *Gini coefficient* [5]. These features are chosen based on major types of statistical measures used in the literature [48], specifically: central tendency (*Mean*, *Median*, and *Mode*), dispersion (*Standard deviation* and *Inter-quartile range*), inequality (*Gini*), and shape (*Skewness* and *Kurtosis*). Recall that in cases when $Size > V_{lead} + V_{trail}$, some data points will not be used in the computation of the features in $F_{complex}$.

For a given set of values $X = \{x_1, x_2, \ldots, x_N\}$ we define the fourteen features as follows:

1. $Size(X) = N$
2. $Min(X) = x_k$ where $x_k \leq x_i \forall 1 \leq i \leq N$
3. $Max(X) = x_k$ where $x_k \geq x_i \forall 1 \leq i \leq N$
4. $Mean(X) = \bar{X} = \frac{\sum_{i=1}^{N} x_i}{N}$
5. $Median(X) = Q_2(X)$ where $Q_2(X)$ is the second quartile of $X$
6. $StdDev(X) = \sqrt{\frac{\sum_{i=1}^{N}(x_i - \bar{X})^2}{N-1}}$
7. $IQR(X) = Q_3(X) - Q_1(X)$ where $Q_1(X)$ and $Q_3(X)$ are the first and third quartiles of $X$ respectively.
8. $Skewness(X) = \frac{\frac{\sum_{i=1}^{N}(x_i - \bar{X})^3}{N}}{\left(\frac{\sum_{i=1}^{N}(x_i - \bar{X})^2}{N-1}\right)^{\frac{3}{2}}}$
9. $Kurtosis(X) = \frac{\frac{\sum_{i=1}^{N}(x_i - \bar{X})^4}{N}}{(StdDev(X))^4} - 3$
10. $Gini(X) = \frac{2\sum_{i=1}^{N} i x_i'}{N \sum_{i=1}^{N} x_i'} - \frac{N+1}{N}$ where $\{x_1', x_2', \ldots, x_N'\}$ is a sorted version of $X$.
11. $Mode(X) = argmax\ |C(x_i)|$ where $1 \leq i \leq N$ and $C(x_i) = \{x \in X$ such that $x = x_i\}$
12. $LongestRunOfZeros(X) = Max(q - p + 1)$ where $1 \leq p \leq q \leq N$ and $x_i = 0\ \forall p \leq i \leq q$
13. $Increasing(X) = \begin{cases} 1 & if\ x_i \leq x_{i+1}\ \forall 1 \leq i < N \\ 0 & otherwise \end{cases}$
14. $Decreasing(X) = \begin{cases} 1 & if\ x_i \geq x_{i+1}\ \forall 1 \leq i < N \\ 0 & otherwise \end{cases}$

Our chosen set of features is by no means the only appropriate one to enable *Substate* profiling. Other potential features will be explored in the future.

*C. ProfileGenerator*

The *ProfileGenerator* creates the *Substate* profiles following the basic high level steps listed below:

1) Given a *Cp* and the *n* test cases that reached it, the corresponding *n* sets of (the fourteen) features are retrieved, one set for each test case.
2) The *n* sets of features (and consequently the *n* test cases) are clustered based on their similarities. *k-means* clustering is used where *k* is a small percentage of *n* and is guaranteed to be ≥ 2. Hereafter, let's assume that *k'* clusters resulted from this step, where $k' \leq k$.
3) Each of the resulting *k'* clusters is deemed to represent a unique *Substate* behavior at the given *Cp* and thus is considered to be a profile element that is covered by an exclusive subset of the *n* test cases.
4) Therefore, at the given *Cp*, *k'* profile elements are identified in addition to the test cases that cover each.
   *Notes: a) It is plausible that a single cluster would result from the above steps due to the n sets of features being identical. b) Test cases that induce one or more NaN values at a given Cp are clustered together, regardless of any other data values they induce; similarly, if the test cases induce one or more Infinity values. The reason is that such test cases share a strongly differentiating trait, in addition, most of the fourteen features cannot be computed given such values.*
5) The above steps are repeated for each *Cp*.
6) As a final step, the *ProfileGenerator* gathers all the profile elements from all the *Cp*'s to build a *Substate* profile for each test case. Recall from step 4) that information about which test cases covered which profile element already exists. The outcome of this step is one file for each test case comprising a sequence of 1's and 0's indicating which profile elements were covered and which were not. It is worth mentioning that the *ProfileGenerator* discards any profile element that is covered by all the test suite.

In the context of test suite reduction, choosing a larger *k* will most likely result in more profile elements to be covered, and more defects to be revealed; however, this occurs at the expense of reduction. In practice, picking the value of *k* really depends on how many test cases the tester is willing to audit; this issue is further discussed in Section V.B (RQ3). In our experiments we varied the value of *k* by setting it to 2 and to the following percentages of *n*: 0.5%, 1%, 1.5%, 2%, 3%, …,10%. The profiles corresponding to these varying values of *k* will be termed: *sstate$_{k@2}$ profiles*, *sstate$_{k@0.5\%}$ profiles*, *sstate$_{k@1\%}$ profiles* etc.

*D. Illustrating Example*

We now walk through an example to illustrate our proposed approach with the help of Figure 2. Consider the BinarytoDecimal program in Section III. The *DataCollector* identifies one *Cp* in main() and eight *Cp*'s in decimal(), specifically:

| | $Cp_1$ | | $Cp_2$ | | $Cp_7$ | | $Cp_8$ | | $Cp_9$ | |
|---|---|---|---|---|---|---|---|---|---|---|
| | $t_1,t_2,t_3,t_4,t_6$ | | $t_1,t_2,t_3,t_4,t_6$ | | $t_1,t_2,t_3,t_4$ | | $t_1,t_2,t_3,t_4$ | | $t_1,t_5,t_6$ | |
| | $t_5$ | | $t_5$ | | $t_5,t_6$ | | $t_5,t_6$ | | $t_2,t_3,t_4$ | |
| | $e_1$ | $e_2$ | $e_3$ | $e_4$ | $e_5$ | $e_6$ | $e_7$ | $e_8$ | $e_9$ | $e_{10}$ |
| $t_1$ | 1 | 0 | 1 | 0 | 1 | 0 | 1 | 0 | 1 | 0 |
| $t_2$ | 1 | 0 | 1 | 0 | 1 | 0 | 1 | 0 | 0 | 1 |
| $t_3$ | 1 | 0 | 1 | 0 | 1 | 0 | 1 | 0 | 0 | 1 |
| $t_4$ | 1 | 0 | 1 | 0 | 1 | 0 | 1 | 0 | 0 | 1 |
| $t_5$ | 0 | 1 | 0 | 1 | 0 | 1 | 0 | 1 | 1 | 0 |
| $t_6$ | 1 | 0 | 1 | 0 | 0 | 1 | 0 | 1 | 1 | 0 |

**Figure 2** – Example overview

- $Cp_1$: at the entry point of main(); initializes the formal parameter args
- $Cp_2$: at the entry point of decimal(); initializes the formal parameter binary
- $Cp_3$: at statement 2; initializes variable decimal to 0
- $Cp_4$: at statement 3, initializes variable i to 0
- $Cp_5$: at statement 3, increments variable i
- $Cp_6$: at statement 4; initializes variable increment to 0
- $Cp_7$: at statement 6; defines variable increment
- $Cp_8$: at statement 7; defines variable decimal
- $Cp_9$: at statement 8; returns the value in decimal

$Cp_3$ through $Cp_6$ are uninteresting as each takes on the same values regardless of the test case being executed. In particular, the data points collected for each of them due to each of the six test cases are as follows: $data(Cp_3)$=<0>, $data(Cp_4)$=<0>, $data(Cp_5)$=<0, 1, 2, 3, 4, 5, 6, 7>, $data(Cp_6)$=<0, 0, 0, 0, 0, 0, 0, 0>. For these *Cp*'s, the extracted features are identical across test cases, and thus *k-means* clustering identifies a single cluster, regardless of the chosen *k*. The single profile element produced for each of these *Cp*'s could be covered by any of the test cases, thus there is no guarantee that any of the failing test cases would be included in the reduced test suite. The *ProfileGenerator* will discard these profile elements in its final step.

The data collected at $Cp_1$ and $Cp_2$ are identical strings. Recall that each data point comprises three values, the length of the string, its richness, and its entropy; therefore, the following tuples are collected at $Cp_1$ and $Cp_2$: $data(t_1)$=<(8, 2, 0.954)>, $data(t_2)$=<(8, 2, 0.954)>, $data(t_3)$=<(8, 2, 0.954)>, $data(t_4)$=<(8, 2, 0.811)>, $data(t_5)$=<(8, 2, 0.543)>, $data(t_6)$=<(8, 2, 0.954)>. Since the length and richness are the same across all six test cases and both *Cp*'s, they will not result in any profile elements. However, applying *k-means* clustering (with *k* = 2) on the features extracted from the entropy values will identify one cluster that includes $t_5$ and a second cluster that includes the other five test cases; thus, resulting in two profile elements at $Cp_1$ and another two at $Cp_2$ (see Figure 2).

$Cp_7$ through $Cp_9$ are most interesting. The values they take on for passing vs. failing test cases are very different. For example, passing test case $t_1$ results in $data(Cp_7)$ = <32, 8, 4, 2, 1>, $data(Cp_8)$ = <0, 0, 32, 32, 40, 44, 46, 47>, and $data(Cp_9)$ = <47>, and failing test case $t_6$ yields $data(Cp_7)$ = <-128, 32, 16, 4, 1>, $data(Cp_8)$ = <-128, -128, -96, -80, -80, -76, -76, -75>, and $data(Cp_9)$ = <-75>. Table 4 shows the computed features at $Cp_7$ for all test cases. Performing *k-means* clustering (with *k* = 2) on these sets of features identifies two clusters that clearly segregate the failing test cases (due to the features annotated with a '*').

| | $t_1$ | $t_2$ | $t_3$ | $t_4$ | $t_5$ | $t_6$ |
|---|---|---|---|---|---|---|
| *Size* | 5 | 5 | 5 | 6 | 7 | 5 |
| *Min\** | 1 | 1 | 4 | 1 | -128 | -128 |
| *Max* | 32 | 64 | 64 | 64 | 64 | 32 |
| *Avg\** | 9.4 | 18.6 | 24.8 | 20.83 | -2.43 | -15 |
| *Med* | 4 | 8 | 16 | 12 | 4 | 4 |
| *Std\** | 12.91 | 25.99 | 24.39 | 23.88 | 59.92 | 64.35 |
| *Iqr* | 18.5 | 37.5 | 42 | 28 | 31 | 87.5 |
| *Gini\** | 0.579 | 0.594 | 0.465 | 0.543 | -10.9 | -1.78 |
| *Skew\** | 0.964 | 0.954 | 0.636 | 0.82 | -1.09 | -0.97 |
| *Kurtosis* | -1.06 | -1.07 | -1.5 | -1.08 | -0.015 | -1.03 |
| *Mode\** | 1 | 1 | 4 | 1 | -128 | -128 |
| *Zeros* | 0 | 0 | 0 | 0 | 0 | 0 |
| *Inc* | 0 | 0 | 0 | 0 | 0 | 0 |
| *Dec\** | 1 | 1 | 1 | 1 | 0 | 0 |

**Table 4** – Features extracted at $Cp_7$

Thus, in order for the two profile elements produced at $Cp_7$ to be covered, either $t_5$ or $t_6$ must be included in the reduced test suite. Similarly, processing the collected data points at $Cp_8$ also results in two clusters that isolate the failing test cases (see Figure 2). However, at $Cp_9$, the collected data at every test is as follows: $data(t_1)$=<47>, $data(t_2)$=<93>, $data(t_3)$=<124>, $data(t_4)$=<125>, $data(t_5)$=<-17>, and $data(t_6)$=<-75>, which results in $t_5$, $t_6$, and $t_1$ to being included in the same cluster. That is the failing tests are clearly segregated in $Cp_7$ and $Cp_8$, but mildly segregated in $Cp_9$.

Since two clusters are identified at each of $Cp_1$ and $Cp_2$, and two clusters at each of $Cp_7$, $Cp_8$, and $Cp_9$, a total of ten profile elements are identified ($e_1, e_2, …, e_{10}$). The *ProfileGenerator* will therefore generate the following profiles for the six test cases: $prof(t_1)$ =<1, 0, 1, 0, 1, 0, 1, 0, 1, 0>, $prof(t_2)$ = $prof(t_3)$ = $prof(t_4)$ = <1, 0, 1, 0, 1, 0, 1, 0, 0, 1>, $prof(t_5)$ = <0, 1, 0, 1, 0, 1, 0, 1, 1, 0>, $prof(t_6)$ = <1, 0, 1, 0, 0, 1, 0, 1, 1, 0>.

To summarize, as shown in Figure 2, $t_5$ is segregated at $Cp_1$ and $Cp_2$, and {$t_5$, $t_6$} are segregated at $Cp_7$ and $Cp_8$. But most importantly, since $e_2$ and $e_4$ are only exercised by $t_5$, a greedy test suite reduction algorithm is guaranteed to include $t_5$ in the reduced test suite and thus the fault will be revealed.

### E. Applications

The *Substate* profiles generated by the *ProfileGenerator* are identical, in regard to format, to the structural profiles described in Section II. Therefore, *Substate Profiling* can be used as the basis for any profile-based technique developed to use structural profiles, including test suite reduction, test suite prioritization, and coverage-based fault localization (CBFL) [26][30][57].

*Substate Profiling* based CBFL identifies failure-correlated profile elements that not only point out suspicious (definition and return) statements but also provide information about the values associated with them, which is very valuable during debugging. Section V presents a study in which an existing greedy test suite reduction algorithm [9][7][29][46] is used in conjunction with *Substate Profiling*. Researchers have also devised variants of this algorithm for test suite prioritization, which could also be used with *Substate Profiling*. For completeness, the greedy TSR algorithm is described next.

Given a program *P*, a test suite *T*, and a set of test requirements *TR* that are covered by *T*. *Test suite reduction* aims at finding *T'*, a minimal subset of *T*, that covers all test requirements in *TR*. The conjecture is that (the smaller) *T'* would be as effective as *T* in revealing defects [46]. Coverage-based *test suite reduction* selects test cases from *T* to include in *T''* in a way that maximizes the proportion of profile elements that are covered. It attempts to cover as many of the elements covered by *T* with as few test cases as possible. A coverage-maximizing subset of a test suite is an instance of the *set-cover problem*, which is NP-complete but which admits a greedy approximation algorithm [9][23]. The greedy algorithm selects the test that covers the largest number of elements not covered by the previously selected tests. This specific approach was termed *basic coverage maximization* in [29][38]. Note that this algorithm might encounter ties; i.e., different tests might each cover the maximal number of elements. In order to break the tie, we randomly select one of the tied test cases, which means that applying the algorithm several times might yield different minimized test suites.

### V. EMPIRICAL EVALUATION

This section presents our comparative empirical evaluation of test suite reduction using *Substate Profiling*.

#### A. Subject Programs

We conducted our experiments using the following Java programs: 1) *NanoXML* releases *r1* through *r5*; 2) the *JTidy* HTML syntax checker and pretty printer *release 3*; 3) the *Xerces* XML parser *release 2.1;* and 4) the seven programs from the *Siemens* benchmark that were translated to Java [1]. The *NanoXML* releases and *Siemens* programs and test suites were download from the SIR repository [13] (*sir.unl.edu*), whereas *JTidy* and *Xerces* were previously used by the authors [33][37][38][36].

*JTidy* and *Xerces* are multi-fault programs. Whereas, each of the *NanoXML* releases and *Siemens* programs is associated with several versions that are seeded with single faults. Since multi-fault programs are more realistic, we created multi-fault versions out of each of the *NanoXML* releases and *Siemens* programs as follows: 1) We randomly selected five of the provided defects such that no two defects involved the same statement. 2) We ran the program using the complete test suite while keeping track of the defects that were triggered within each test case. 3) We discarded all test cases that triggered more than one defect; as such, the final number of defects we considered varied between 1 and 5 as some defects were never triggered alone. 4) We randomly reduced the number of test cases to arrive at a

| | LOC | #def | |T| | # Failures | | | | | | | |
|---|---|---|---|---|---|---|---|---|---|---|---|
| | | | | Total | def1 | def2 | def3 | def4 | def5 | def6 | def7 | def8 |
| *JTidy* | 9.1K | 8 | 1000 | 47 | 5 | 6 | 16 | 9 | 1 | 5 | 4 | 1 |
| *Xerces* | 52.5K | 6 | 1519 | 24 | 3 | 3 | 5 | 4 | 4 | 5 | - | - |
| *Nano r1* | 4.3K | 2 | 212 | 30 | 20 | 10 | - | - | - | - | - | - |
| *Nano r2* | 5.8K | 3 | 177 | 20 | 15 | 1 | 4 | - | - | - | - | - |
| *Nano r3* | 7.2K | 4 | 215 | 29 | 8 | 16 | 4 | 1 | - | - | - | - |
| *Nano r5* | 7.5K | 1 | 185 | 50 | 50 | - | - | - | - | - | - | - |
| *replace* | 554 | 4 | 1158 | 158 | 8 | 50 | 50 | 50 | - | - | - | - |
| *tot_info* | 494 | 5 | 423 | 105 | 6 | 9 | 33 | 50 | 7 | - | - | - |
| *print_tok* | 536 | 3 | 1121 | 121 | 50 | 50 | 21 | - | - | - | - | - |
| *print_tok2* | 387 | 4 | 1200 | 200 | 50 | 50 | 50 | 50 | - | - | - | - |
| *schedule* | 425 | 4 | 1000 | 173 | 23 | 50 | 50 | 50 | - | - | - | - |
| *schedule2* | 766 | 4 | 1172 | 172 | 50 | 50 | 34 | 38 | - | - | - | - |
| *tcas* | 136 | 5 | 931 | 157 | 9 | 45 | 3 | 50 | 50 | - | - | - |

**Table 5** – Subjects, test suites, defects, and failures.

|  | **BB** | | **BBE** | | **DUP** | | **ALL** | | **k@2** | | **k@.5%** | | **k@1%** | | **k@1.5%** | | **k@2%** | | **k@3%** | | **k@4%** | | **k@5%** | | **k@6%** | | **k@7%** | | **Better** | |
|---|---|---|---|---|---|---|---|---|---|---|---|---|---|---|---|---|---|---|---|---|---|---|---|---|---|---|---|---|---|
|  | rd% | df% | rd% | df% | rd% | df% | rd% | df% | rd% | df% | rd% | df% | rd% | df% | rd% | df% | rd% | df% | rd% | df% | rd% | df% | rd% | df% | rd% | df% | rd% | df% | State | Struct |
| *JTidy* | 94 | 60 | 93 | 71 | 90 | 87 | 89 | 87 | 92 | 76 | 90 | 80 | 84 | 86 | 79 | 93 | 75 | 100 | 67 | 100 | 60 | 100 | 55 | 100 | 51 | 100 | 47 | 100 | ✓ |  |
| *Xerces* | 89 | 65 | 88 | 58 | 84 | 66 | 83 | 66 | 85 | 71 | 84 | 86 | 80 | 100 | 76 | 100 | 72 | 100 | 67 | 100 | 59 | 100 | 56 | 100 | 52 | 100 | 49 | 100 | ✓ |  |
| *Nano r1* | 88 | 100 | 87 | 100 | 88 | 100 | 84 | 100 | 86 | 100 | 86 | 100 | 87 | 100 | 86 | 100 | 83 | 100 | 77 | 100 | 72 | 100 | 68 | 100 | 64 | 100 | 62 | 100 | - | - |
| *Nano r2* | 85 | 100 | 85 | 100 | 84 | 100 | 82 | 100 | 83 | 100 | 83 | 100 | 83 | 100 | 83 | 100 | 83 | 100 | 76 | 100 | 72 | 100 | 68 | 100 | 65 | 100 | 63 | 100 | - | - |
| *Nano r3* | 86 | 60 | 86 | 67 | 86 | 62 | 84 | 65 | 82 | 100 | 82 | 100 | 82 | 100 | 82 | 100 | 79 | 100 | 73 | 100 | 67 | 100 | 64 | 100 | 61 | 100 | 58 | 100 | ✓ |  |
| *Nano r5* | 87 | 100 | 85 | 100 | 85 | 100 | 83 | 100 | 83 | 100 | 83 | 100 | 83 | 100 | 83 | 100 | 82 | 100 | 81 | 100 | 78 | 100 | 75 | 100 | 72 | 100 | 69 | 100 | - | - |
| *replace* | 99 | 62 | 99 | 75 | 99 | 68 | 99 | 75 | 99 | 81 | 98 | 100 | 96 | 100 | 94 | 100 | 92 | 100 | 89 | 100 | 85 | 100 | 82 | 100 | 80 | 100 | 78 | 100 | ✓ |  |
| *tot_info* | 97 | 45 | 97 | 44 | 98 | 60 | 97 | 60 | 96 | 65 | 96 | 75 | 95 | 64 | 93 | 65 | 90 | 80 | 83 | 95 | 79 | 90 | 74 | 95 | 70 | 89 | 65 | 90 | ✓ |  |
| *print_tok* | 99 | 100 | 99 | 100 | 94 | 100 | 94 | 100 | 98 | 100 | 97 | 100 | 95 | 100 | 92 | 100 | 90 | 100 | 86 | 100 | 82 | 100 | 79 | 100 | 76 | 100 | 73 | 100 | - | - |
| *print_tok2* | 99 | 75 | 99 | 100 | 99 | 100 | 99 | 100 | 99 | 100 | 99 | 100 | 95 | 100 | 93 | 100 | 91 | 100 | 88 | 100 | 84 | 100 | 79 | 100 | 76 | 100 | 73 | 100 | - | - |
| *schedule* | 99 | 50 | 99 | 50 | 98 | 75 | 97 | 75 | 98 | 68 | 98 | 75 | 95 | 81 | 93 | 87 | 91 | 87 | 86 | 81 | 82 | 93 | 77 | 87 | 74 | 100 | 70 | 100 | ✓ |  |
| *schedule2* | 99 | 100 | 99 | 100 | 99 | 100 | 99 | 100 | 98 | 100 | 97 | 100 | 94 | 100 | 91 | 100 | 88 | 100 | 82 | 100 | 76 | 100 | 72 | 100 | 66 | 100 | 62 | 100 | - | - |
| *tcas* | 99 | 55 | 98 | 65 | 99 | 35 | 98 | 50 | 99 | 70 | 98 | 75 | 97 | 80 | 96 | 85 | 95 | 95 | 93 | 100 | 91 | 100 | 89 | 100 | 87 | 100 | 84 | 100 | ✓ |  |

**Table 6** – $rd_\%$ and $df_\%$ computed (given all failures)

|  | **BB** | | **BBE** | | **DUP** | | **ALL** | | **k@2** | | **k@.5%** | | **k@1%** | | **k@1.5%** | | **k@2%** | | **k@3%** | | **k@4%** | | **k@5%** | | **k@6%** | | **k@7%** | | **Better** | |
|---|---|---|---|---|---|---|---|---|---|---|---|---|---|---|---|---|---|---|---|---|---|---|---|---|---|---|---|---|---|
|  | rd% | df% | rd% | df% | rd% | df% | rd% | df% | rd% | df% | rd% | df% | rd% | df% | rd% | df% | rd% | df% | rd% | df% | rd% | df% | rd% | df% | rd% | df% | rd% | df% | State | Struct |
| *JTidy* | 94 | 43 | 93 | 62 | 90 | 75 | 89 | 77 | 92 | 67 | 90 | 73 | 85 | 81 | 80 | 89 | 75 | 89 | 67 | 93 | 61 | 94 | 56 | 99 | 51 | 100 | 48 | 99 | ✓ |  |
| *Xerces* | 89 | 47 | 88 | 50 | 84 | 52 | 83 | 63 | 85 | 47 | 84 | 61 | 80 | 85 | 76 | 89 | 73 | 92 | 67 | 97 | 62 | 99 | 59 | 100 | 56 | 100 | 53 | 100 | ✓ |  |
| *Nano r1* | 86 | 55 | 85 | 55 | 86 | 55 | 83 | 55 | 85 | 60 | 86 | 65 | 85 | 65 | 85 | 70 | 81 | 85 | 75 | 100 | 70 | 100 | 65 | 100 | 63 | 100 | 60 | 100 | ✓ |  |
| *Nano r2* | 84 | 100 | 84 | 100 | 83 | 100 | 81 | 100 | 82 | 100 | 83 | 100 | 82 | 100 | 82 | 100 | 82 | 100 | 75 | 100 | 71 | 100 | 66 | 100 | 64 | 100 | 60 | 100 | - | - |
| *Nano r3* | 87 | 52 | 87 | 52 | 88 | 60 | 85 | 50 | 83 | 92 | 83 | 87 | 83 | 97 | 83 | 95 | 80 | 97 | 74 | 100 | 68 | 85 | 64 | 97 | 62 | 92 | 59 | 95 | ✓ |  |
| *Nano r5* | 82 | 100 | 82 | 100 | 80 | 100 | 77 | 100 | 79 | 100 | 79 | 100 | 79 | 100 | 79 | 100 | 78 | 100 | 77 | 100 | 74 | 100 | 72 | 100 | 68 | 100 | 65 | 100 | - | - |
| *replace* | 99 | 25 | 99 | 50 | 99 | 25 | 98 | 50 | 98 | 25 | 97 | 31 | 95 | 37 | 93 | 37 | 90 | 37 | 87 | 56 | 83 | 62 | 79 | 68 | 77 | 81 | 74 | 93 | ✓ |  |
| *tot_info* | 97 | 25 | 97 | 25 | 96 | 44 | 97 | 40 | 96 | 60 | 95 | 50 | 95 | 55 | 94 | 50 | 91 | 55 | 89 | 55 | 82 | 65 | 77 | 65 | 71 | 60 | 69 | 70 | 65 | 75 | ✓ |  |
| *print_tok* | 99 | 100 | 99 | 100 | 93 | 66 | 93 | 100 | 98 | 83 | 97 | 83 | 94 | 91 | 92 | 83 | 90 | 91 | 86 | 91 | 83 | 100 | 80 | 100 | 77 | 100 | 74 | 100 |  | ✓ |
| *print_tok2* | 99 | 75 | 99 | 100 | 99 | 75 | 99 | 100 | 99 | 75 | 97 | 75 | 95 | 81 | 93 | 81 | 90 | 81 | 86 | 81 | 83 | 93 | 78 | 93 | 75 | 87 | 72 | 87 |  | ✓ |
| *schedule* | 99 | 50 | 98 | 50 | 98 | 75 | 98 | 75 | 98 | 56 | 97 | 43 | 94 | 75 | 92 | 75 | 89 | 81 | 84 | 81 | 79 | 93 | 75 | 93 | 71 | 100 | 67 | 93 | ✓ |  |
| *schedule2* | 99 | 56 | 99 | 56 | 99 | 56 | 99 | 68 | 99 | 68 | 97 | 81 | 94 | 93 | 89 | 100 | 86 | 100 | 79 | 100 | 74 | 100 | 69 | 100 | 63 | 100 | 59 | 100 | ✓ |  |
| *tcas* | 99 | 30 | 98 | 25 | 99 | 5 | 98 | 25 | 99 | 35 | 98 | 35 | 96 | 30 | 95 | 35 | 94 | 35 | 92 | 35 | 89 | 35 | 87 | 35 | 84 | 40 | 81 | 45 | ✓ |  |

**Table 7** – $rd_\%$ and $df_\%$ computed given a single failure per defect

maximum of 50 failing tests per defect and a maximum of 1000 passing tests.

Since the aim of our study is to assess the effectiveness of the reduced test suites at revealing faults, we discarded the defects that exhibited no failures, and discarded *NanoXML r4* as no defects were associated with it. Also, due to a limitation in the current implementation of the *DataCollector*, some exceptions behaved in a manner that prevented the proper collection of state data; this led us to discard one defect in *NanoXML r1* and two defects in *NanoXML r5*. Finally, in order to factor out the negative impact of coincidental correctness on the accuracy of our comparative study [24][33], we considered a test case to be failing not only when it produces an unexpected output but when an infection is detected right after the defect is exercised (regardless of the output). This same approach was followed in several of our previous studies [33][37][38][36].

Table 5 provides information about our subject programs: a) code sizes; b) number of defects; c) test suite sizes; d) total number of failures; and e) number of failures per defect. It is worth noting that the execution traces of *JTidy* and *Xerces* are considerably longer than for the other programs; this is not apparent in Table 5 but it becomes clear in Table 10, which shows the average time cost of collecting execution profiles.

### B.  Results and Observations

For each subject, we performed test suite reduction using structural profiles of the following types: *BB*, *BBE*, *DUP*, and *ALL*. Since the greedy reduction algorithm is not deterministic, we applied it 100 times for each subject and computed the average sizes of the reduced test suites and the average numbers of the revealed defects. Accordingly, Table 6 reports for each profile type and program, the average percentage reduction in test suite size ($rd_\%$), and the average percentage of revealed defects ($df_\%$). In the case of *Substate* profiles, we performed reduction for thirteen values of *k*, starting with *k*=2 then setting it to the following percentages of *n*: 0.5%, 1%, 1.5%, 2%, 3%, …, 10%. Table 6 shows all the $rd_\%$ and $df_\%$ values, except for those corresponding to *k@8%* through *k@10%*, which were omitted for space limitations. The two right most columns in Table 6 indicate which of the approaches performed better according to the following: a) *ALL* is considered as a representative of structural profiling; b) the approach exhibiting the higher $df_\%$ is deemed better, as long as $rd_\%$ is not diminished by more than 20%. The cells in Table 6 reflecting the better results are shaded in grey. Next, we present our results by answering relevant research questions.

**RQ1:** *How does Substate Profiling perform relative to Structural Profiling?* – Table 6 allows us to make the following observations:
1) *JTidy*: in regard to defect detection, $sstate_{k@2}$ profiles performed better than *BB* and *BBE*, and $sstate_{k@1.5\%}$ performed better than *DUP* and *ALL*. In addition, *Substate* profiling covered 100% of the defects starting with *k@2%*. Regarding test suite reduction, the $rd_\%$ for *k@2* is comparable to that of *BB* and *BBE*, however, the $rd_\%$ for *k@1.5%* and *k@2%* are considerably lower than that for *DUP* and *ALL*. Only the cells corresponding to *k@1.5%* and *k@2%* are shaded (considered better performers than *ALL*) since their $df_\%$ is higher than the $df_\%$ for *ALL*, and the difference between their $rd_\%$ and that of *ALL* is less than

20%. Therefore, for *JTidy*, *Substate* profiling performed better than structural profiling, as indicated by the check mark in the right most columns of Table 6.
2) *Xerces*: in regard to defect detection, *Substate* profiling performed better than any of the structural profiles starting with *k@2*, and covered 100% of the defects starting with *k@1%*. Noting that for *k@2* and *k@1%* the $rd_\%$ is somewhat comparable to that for *ALL*. The cells corresponding to *k@4%* through *k@7%* are not shaded since they reflect $rd_\%$ values that are relatively very low.
3) *NanoXML r1, r2, and r5*: in these three cases *Substate* profiling and structural profiling performed equally well. Compared to *ALL*, the $rd_\%$ values are almost equal for small *k* values but much smaller for large *k* values, which is expected.
4) *NanoXML r3*: *Substate* profiling covered 100% of the defects starting with *k@2%*, whereas none of the strutural profiles performed well.
5) *replace*: *k@2* performed better than any of the structural profiles, and starting with *k@.5%*, 100% of the defects are covered.
6) *tot_info*: *k@2* performed better than any of the structural profiles.
7) *print_token, print_token2*, and *schedule2*: in these three cases, *Substate* profiling and structural profiling performed equally well.
8) *schedule*: *k@1%* performed better than the structural profiles, and 100% of the defects are covered starting with *k@6%*.
9) *tcas*: similarly, *k@2* performed better than any of the structural profiles, and starting with *k@3%*, 100% of the defects are covered.

In summary, *Substate* profiling performed better for seven subjects, and equally well for six subjects. It should be noted however that our criteria for ranking the two approaches gives more weight to defect coverage than to test suite reduction.

**RQ2:** *What impact does the number of failures have on the effectiveness of the techniques?* – As shown in Table 5, some of the defects have a large number of failures associated with them, which is not typical in a realistic setting. For example, *def3* of *JTidy* induces *16* failures, and *def2* of *replace* induces 50 failures. To better assess the effectiveness of *Substate* profiling in a realistic environment, we modified the experiments described in RQ1 by considering a single failure per defect. Specifically, we randomly considered only one failure from each defect, then applied greedy reduction. This was repeated *n* times such as *n* = 10× (original number of failures). Table 7 reports the corresponding averages and allows us to make the following observations:
1) *NanoXML r1* and *schedule2*: for these two subjects, there was tie between the two approaches when considering all failures. However, when a single failure was considered, *Substate* profiling perfomed better.
2) *print_token* and *print_token2*: for these cases, there was also a tie between the two approaches when considering all failures, but when a single failure was considered, *structural* profiling perfomed better. We examined the defects in these two programs and realized that both *print_token* and *print_token2* have faulty conditional statements, which are expected to be detected by *structural* profiling, but not necessarily by *Substate* profiling. Since the latter considers only definition and return statements.
3) The outcome of the two approaches was unchanged for the remaining nine subjects.

Using single failures, *Substate* profiling performed better for nine subjects, worse for two subjects, and equally well for two subjects. Clearly, using a single failure is the more realistic approach to evaluate our proposed technique. In our study, using all failures yielded different results than those for single failures, however, the conclusions are to a great extent the same.

**RQ3:** *So what value of k should a tester use?* – Table 6 suggests that *k@2%* is optimal for *JTidy*, *k@1.5%* is optimal for *Xerces*, and *k@2* is optimal for *NanoXML r3*. Therefore, there does not seem to be an optimal value of *k* that testers should use for all applications. However, it should be reasonable to assume that a larger *k* is likely to yield more defect coverage at the cost of diminished test suite reduction. In addition, testers could experiment with different values of *k* on prior releases of their products; thus conjecturing that values of *k* that worked well in prior releases would also yield good results in the current release.

**RQ4:** *Are Substate profiles complementary to structural profiles?* - Our goal here is to explore whether combining structural profiles with *Substate* profiles generated using small values of *k* (specifically, *k@2*, *k@.5%*, *k@1%* and *k@2%*), would yield better results than when each approach is used separately. In this context, a better result is one that: 1) reveals more defects (higher $df_\%$) than both separate approaches; and 2) in case the same number of defects is revealed (i.e., a tie), the better result is the one with the fewer selected test cases (higher $rd_\%$). Table 8 shows the results when combining *BB*, *BBE*, *DUP*, and *ALL* profiles, each with $sstate_{k@2}$, $sstate_{k@.5\%}$, $sstate_{k@1\%}$, and $sstate_{k@2\%}$ profiles. For example, column "*BBE + .5%*", shows the results for combining *BBE* profiles with $sstate_{k@.5\%}$ profiles. The highlighted cells correspond to the cases when the combined techniques performed better than both of the original techniques. For example, the "*BBE + .5%*" result for *Xerces* is highlighted since "*78/92*" is better than its *BBE* and $sstate_{k@.5\%}$

|  | Collect (a) | | Processing + Reduction (b) | | Total Cost (c) | |
| --- | --- | --- | --- | --- | --- | --- |
|  | ALL | sstate | ALL | sstate | ALL | sstate |
| *JTidy* | 30 | 31 | 315 | 3755 | 30315 | 34755 |
| *Xerces* | 15 | 6 | 978 | 829 | 23748 | 9937 |
| *Nano1* | 0.5 | 0.4 | 10 | 5 | 108 | 95 |
| *Nano2* | 0.5 | 0.4 | 8 | 5 | 95 | 81 |
| *Nano3* | 0.5 | 0.4 | 14 | 7 | 122 | 104 |
| *Nano5* | 0.5 | 0.5 | 12 | 5 | 103 | 90 |
| *replace* | 0.7 | 0.6 | 13 | 12 | 813 | 665 |
| *tot_info* | 0.7 | 0.6 | 5 | 17 | 301 | 291 |
| *print_tok* | 0.7 | 0.6 | 17 | 62 | 780 | 742 |
| *print_tok2* | 0.7 | 0.6 | 9 | 25 | 818 | 746 |
| *schedule* | 0.7 | 0.6 | 10 | 9 | 694 | 614 |
| *schedule2* | 0.7 | 0.7 | 17 | 44 | 841 | 847 |
| *tcas* | 0.7 | 0.5 | 3 | 4 | 624 | 497 |

**Table 10** – Cost analysis (in secs): a) Avg time for collecting a single profile. b) Time for processing the profiles and for minimization (single replication)

|  | BB+2 | | BB+.5% | | BB+1% | | BB+2% | | BBE+2 | | BBE+.5% | | BBE+1% | | BBE+2% | | DUP+2 | | DUP+.5% | | DUP+1% | | DUP+2% | | ALL+2 | | ALL+.5% | | ALL+1% | | ALL+2% | |
|---|---|---|---|---|---|---|---|---|---|---|---|---|---|---|---|---|---|---|---|---|---|---|---|---|---|---|---|---|---|---|---|---|
|  | rd% | df% | rd% | df% | rd% | df% | rd% | df% | rd% | df% | rd% | df% | rd% | df% | rd% | df% | rd% | df% | rd% | df% | rd% | df% | rd% | df% | rd% | df% | rd% | df% | rd% | df% | rd% | df% |
| *JTidy* | 90 | 79 | 88 | 81 | 83 | 86 | 74 | 98 | 89 | 80 | 87 | 78 | 82 | 86 | 73 | 100 | 87 | 90 | 85 | 90 | 80 | 93 | 72 | 100 | 86 | 92 | 85 | 92 | 80 | 95 | 72 | 100 |
| *Xerces* | 80 | 82 | 79 | 83 | 75 | 100 | 69 | 100 | 80 | 85 | 78 | 92 | 75 | 100 | 69 | 100 | 77 | 82 | 76 | 85 | 73 | 100 | 67 | 100 | 77 | 84 | 75 | 87 | 72 | 100 | 67 | 100 |
| *Nano r1* | 79 | 100 | 80 | 100 | 79 | 100 | 76 | 100 | 79 | 100 | 79 | 100 | 79 | 100 | 75 | 100 | 79 | 100 | 79 | 100 | 80 | 100 | 76 | 100 | 77 | 100 | 77 | 100 | 77 | 100 | 74 | 100 |
| *Nano r2* | 76 | 100 | 76 | 100 | 75 | 100 | 75 | 100 | 76 | 100 | 76 | 100 | 76 | 100 | 75 | 100 | 76 | 100 | 76 | 100 | 76 | 100 | 75 | 100 | 74 | 100 | 74 | 100 | 74 | 100 | 73 | 100 |
| *Nano r3* | 76 | 100 | 76 | 100 | 76 | 100 | 74 | 100 | 75 | 100 | 76 | 100 | 76 | 100 | 73 | 100 | 77 | 100 | 77 | 100 | 77 | 100 | 75 | 100 | 74 | 100 | 74 | 100 | 74 | 100 | 71 | 100 |
| *Nano r5* | 78 | 100 | 78 | 100 | 78 | 100 | 78 | 100 | 76 | 100 | 76 | 100 | 76 | 100 | 76 | 100 | 76 | 100 | 76 | 100 | 76 | 100 | 76 | 100 | 75 | 100 | 75 | 100 | 75 | 100 | 75 | 100 |
| *replace* | 98 | 81 | 97 | 100 | 95 | 100 | 92 | 100 | 98 | 93 | 97 | 100 | 95 | 100 | 92 | 100 | 98 | 75 | 97 | 100 | 95 | 100 | 92 | 100 | 98 | 81 | 97 | 100 | 95 | 100 | 91 | 100 |
| *tot_info* | 94 | 75 | 94 | 60 | 93 | 65 | 88 | 85 | 94 | 60 | 94 | 65 | 93 | 80 | 88 | 85 | 94 | 70 | 94 | 65 | 93 | 70 | 89 | 75 | 93 | 65 | 94 | 70 | 93 | 65 | 88 | 80 |
| *print_tok* | 98 | 100 | 97 | 100 | 95 | 100 | 90 | 100 | 98 | 100 | 97 | 100 | 95 | 100 | 90 | 100 | 93 | 100 | 92 | 100 | 91 | 100 | 86 | 100 | 93 | 100 | 92 | 100 | 90 | 100 | 86 | 100 |
| *print_tok2* | 98 | 100 | 97 | 100 | 95 | 100 | 91 | 100 | 98 | 100 | 97 | 100 | 95 | 100 | 91 | 100 | 98 | 100 | 97 | 100 | 95 | 100 | 91 | 100 | 98 | 100 | 97 | 100 | 95 | 100 | 91 | 100 |
| *schedule* | 98 | 75 | 97 | 81 | 95 | 75 | 90 | 93 | 98 | 75 | 97 | 81 | 95 | 81 | 90 | 75 | 97 | 100 | 96 | 100 | 94 | 93 | 89 | 100 | 97 | 100 | 96 | 100 | 94 | 93 | 89 | 93 |
| *schedule2* | 98 | 100 | 97 | 100 | 94 | 100 | 87 | 100 | 98 | 100 | 97 | 100 | 94 | 100 | 88 | 100 | 98 | 100 | 97 | 100 | 94 | 100 | 88 | 100 | 98 | 100 | 97 | 100 | 94 | 100 | 88 | 100 |
| *tcas* | 98 | 80 | 98 | 85 | 96 | 85 | 94 | 100 | 98 | 80 | 97 | 80 | 96 | 80 | 94 | 100 | 98 | 80 | 98 | 85 | 97 | 75 | 95 | 90 | 98 | 85 | 97 | 85 | 96 | 95 | 94 | 95 |

**Table 8** – Complementary $rd_\%$ and $df_\%$ (given all failures)

|  | BB+2 | | BB+.5% | | BB+1% | | BB+2% | | BBE+2 | | BBE+.5% | | BBE+1% | | BBE+2% | | DUP+2 | | DUP+.5% | | DUP+1% | | DUP+2% | | ALL+2 | | ALL+.5% | | ALL+1% | | ALL+2% | |
|---|---|---|---|---|---|---|---|---|---|---|---|---|---|---|---|---|---|---|---|---|---|---|---|---|---|---|---|---|---|---|---|---|
|  | rd% | df% | rd% | df% | rd% | df% | rd% | df% | rd% | df% | rd% | df% | rd% | df% | rd% | df% | rd% | df% | rd% | df% | rd% | df% | rd% | df% | rd% | df% | rd% | df% | rd% | df% | rd% | df% |
| *JTidy* | 90 | 76 | 88 | 77 | 83 | 84 | 74 | 91 | 89 | 80 | 87 | 79 | 82 | 84 | 74 | 91 | 86 | 90 | 85 | 93 | 80 | 95 | 72 | 98 | 86 | 87 | 85 | 92 | 80 | 93 | 72 | 97 |
| *Xerces* | 80 | 63 | 79 | 74 | 75 | 89 | 69 | 94 | 80 | 64 | 78 | 72 | 75 | 88 | 69 | 93 | 77 | 70 | 76 | 79 | 73 | 93 | 67 | 96 | 76 | 69 | 75 | 78 | 72 | 91 | 67 | 97 |
| *Nano r1* | 78 | 65 | 78 | 70 | 78 | 70 | 74 | 85 | 77 | 65 | 77 | 70 | 77 | 65 | 72 | 75 | 78 | 65 | 78 | 70 | 78 | 75 | 74 | 80 | 74 | 75 | 75 | 65 | 75 | 65 | 71 | 75 |
| *Nano r2* | 75 | 100 | 75 | 100 | 75 | 100 | 74 | 100 | 75 | 100 | 74 | 100 | 74 | 100 | 74 | 100 | 75 | 100 | 74 | 100 | 74 | 100 | 74 | 100 | 73 | 100 | 72 | 100 | 72 | 100 | 71 | 100 |
| *Nano r3* | 77 | 100 | 77 | 95 | 77 | 95 | 74 | 97 | 77 | 95 | 77 | 97 | 76 | 92 | 73 | 100 | 78 | 100 | 78 | 95 | 78 | 100 | 75 | 95 | 75 | 95 | 75 | 92 | 75 | 92 | 72 | 97 |
| *Nano r5* | 72 | 100 | 72 | 100 | 73 | 100 | 72 | 100 | 72 | 100 | 72 | 100 | 72 | 100 | 71 | 100 | 72 | 100 | 71 | 100 | 71 | 100 | 71 | 100 | 70 | 100 | 69 | 100 | 70 | 100 | 69 | 100 |
| *replace* | 98 | 25 | 97 | 25 | 94 | 37 | 90 | 43 | 98 | 50 | 97 | 50 | 94 | 62 | 90 | 68 | 98 | 25 | 97 | 25 | 94 | 50 | 90 | 50 | 97 | 50 | 96 | 50 | 94 | 56 | 90 | 62 |
| *tot_info* | 93 | 50 | 93 | 50 | 92 | 55 | 87 | 55 | 93 | 60 | 93 | 60 | 92 | 60 | 86 | 65 | 94 | 44 | 94 | 50 | 92 | 50 | 87 | 55 | 93 | 60 | 93 | 60 | 92 | 65 | 86 | 60 |
| *print_tok* | 98 | 100 | 97 | 100 | 94 | 100 | 90 | 100 | 98 | 100 | 97 | 100 | 94 | 100 | 90 | 100 | 93 | 83 | 92 | 74 | 90 | 91 | 86 | 91 | 93 | 100 | 92 | 100 | 90 | 100 | 86 | 100 |
| *print_tok2* | 98 | 75 | 97 | 87 | 95 | 81 | 90 | 81 | 98 | 100 | 97 | 100 | 94 | 100 | 90 | 100 | 98 | 75 | 97 | 81 | 95 | 81 | 90 | 81 | 98 | 100 | 97 | 100 | 95 | 100 | 91 | 100 |
| *schedule* | 98 | 56 | 97 | 50 | 94 | 75 | 89 | 75 | 97 | 50 | 97 | 50 | 94 | 68 | 89 | 81 | 97 | 75 | 96 | 75 | 94 | 87 | 88 | 87 | 97 | 81 | 96 | 81 | 93 | 81 | 88 | 100 |
| *schedule2* | 98 | 68 | 97 | 81 | 93 | 93 | 86 | 100 | 98 | 68 | 97 | 87 | 93 | 87 | 86 | 93 | 98 | 81 | 96 | 81 | 93 | 93 | 85 | 100 | 98 | 75 | 96 | 81 | 93 | 93 | 86 | 100 |
| *tcas* | 98 | 45 | 97 | 50 | 96 | 44 | 94 | 55 | 98 | 50 | 97 | 44 | 95 | 44 | 93 | 50 | 98 | 35 | 97 | 35 | 96 | 25 | 94 | 35 | 98 | 50 | 97 | 50 | 95 | 50 | 93 | 44 |

**Table 9** – Complementary $rd_\%$ and $df_\%$ computed given a single failure per defect

results, which are "*88/58*" and "*84/86*", respectively. In this case, the result is considered better as it revealed more defects than when using both *BBE* and $sstate_{k@.5\%}$. Now consider the "*BBE + 1%*" result for *JTidy*, namely, "*82/86*". This result is not highlighted since it is not deemed better than its corresponding $sstate_{k@1\%}$ result, namely, "*84/86*". In all, 53 out of the 208 results were better than the results generated by either of the original techniques. Table 9 shows the equivalent results for the case when a single failure per defect is considered. For that, the improved results were observed in 87 out of the 208 cases. These findings suggest that there is a considerable chance that:

1) Combining *Substate* profiles with structural profiles would be beneficial.
2) *Substate* profiles are complementary to structural profiles.

### B. Cost Analysis

Regardless of the profile type used, coverage-based test suite reduction involves three main steps: 1) collecting the profiles; 2) processing and consolidating the profiles; and 3) greedy reduction. Table 10 shows for *ALL* and *Substate* profiling: a) the average times for collecting a single profile; b) the times for processing the profiles, and for executing a single iteration of the reduction step; and c) the total costs incurred, which take into consideration the execution of the complete test suites. Based on the reported total costs, *Substate* profiling is less costly except for the case of *JTidy*. However, the cost difference is minor in most cases. More importantly, all measured times are likely to be insignificant in cases when manual auditing is required and/or when the cost of test execution is hindering, as discussed in Section I.

### C. Threats to Validity

Our experiments involved a limited number of programs and defects which requires us to conduct more extensive experiments in the future that include more programs that are larger and are drawn from different domains.

Our approach enables the tester to choose a value of *k* that is proportional to the resources available for auditing. However, we believe that the tester needs more information about the impact of a given value of *k* in practice. In Section V.B.RQ3, we suggested that prior releases of the software product could be leveraged to determine an effective value of *k*; however, more needs to be done in relation to this issue.

Finally, our study uses *BB*, *BBE*, *DUP*, and *ALL* as representative structural profiling elements; whereas, other suitable structural elements do exist. Actually, the work presented in [38] shows that slice pair profiling is very effective in regard to defect detection. However, it greatly suffers in regard to test suite size reduction, in addition, the cost of collecting and processing slice pair profiles is hindering, which is not the case in *Substate Profiling*.

## VI. RELATED WORK

The focus of this section is on work related to state profiling. For work related to test suite reduction, the reader can refer to [44][45][46][51][52][53] for early work, to [31][18][49][12] for more recent work, and to the comprehensive survey provided by Yoo and Harman [56]. The authors' previous contribution to test suite reduction can be found in [38][35].

Zimmerman and Zeller [59] modeled memory as a graph that captures program state. Values are represented as vertices and references between values as edges. One approach for state profiling would be to capture memory graphs following the execution of definition statements, another would be to capture them at the end of program execution. However, the former is clearly infeasible, and the latter will likely miss states relevant to failure. Our approach attempts to strike the right balance between these two extreme forms of state profiling.

Xie and Notkin [55] profiled the values of global variables and function parameters at the entry and exit of functions in order to collect and compare state profiles, which they termed "value spectra", for the purpose of regression testing. The goal of the comparison is to detect internal behavior deviations between two program versions even when program outputs are the same. We ran part of our experiments (using *JTidy* and *Xerces*) while only considering static variables and formal parameters. We observed a measurable improvement in regard to reduction, but a major deterioration with respect to the number of bugs revealed.

Xie *et al.* [54] minimized unit test suites by collecting the state of objects at the entry of unit test functions; a test initiated with a previously observed state is deemed redundant. Their proposed framework, Rostra, was successfully used to minimize test suites generated by test case generation tools such as JTest and JCrasher. *Substate Profiling* has a wider scope, in regard to applications and the level of testing.

Gyori *et al.* proposed PolDet [19], a technique that detects polluting unit tests; i.e., tests that might cause subsequently executed tests to fail as a result of erroneously modifying data they share with them. Given a unit test *t*, PolDet captures the program's heap-state before *t*'s setup and teardown code. The two captured states are compared in order to determine whether or not their differences are potentially harmful for subsequently executing tests. For example, *t* is deemed non-polluting if the captured states are isomorphic, even if the recorded objects field values are different. Also, private fields and local variables are not considered when capturing the states, since they cannot be accessed by other tests and thus cannot be harmful. PolDet's approach to capturing and comparing states appears to be relevant to our work, however it is not useful for general state profiling since it operates at a level of abstraction that is too high for its purposes. Specifically, state behaviors occurring during a test run might not always be reflected in the final captured state, and uncaptured state behaviors involving local variables and private fields might be relevant to failure.

Francis [16] collected object states to enable test suite reduction. However, their empirical results were not as favorable as ours. The *JTidy* subject program we used is identical to the one they used. As presented in [16], their approach revealed 71% of the defects with an $rd_\%$ of 77, whereas *Substate Profiling* revealed roughly 96% for the same $rd_\%$, and 100% for an $rd_\%$ of 75 (see the results for $k@1.5\%$ and $k@2\%$ in Table 6). We also used the same version of *Xerces* they used, but with a slightly different set of defects. Their approach revealed all the defects with an $rd_\%$ of 50, whereas ours revealed all the defects with an $rd_\%$ of 80. We believe that object-state profiling [16] does not perform as well because it might miss many object states that could be relevant to failure. This might happen for two reasons: 1) the object states are only collected at the exit points of functions; and 2) only samples of them are actually collected [16].

Elbaum *et al.* [15] presented a framework for creating and replaying tests that are hybrid of unit and system tests, which they termed *Differential Unit Tests*. Their approach involves analyzing program states acquired before and after the execution of a given unit test. Since they recognize that recording raw program states is impractical, they adopted several strategies to approximate them. For example, they considered a single representative of each equivalence class of program states. They also considered only the values of reference fields and discarded scalar fields, which would maintain the heap shape of a program state. The framework was subsequently extended [27] to create *Aggregated Differential Unit Tests*. In future work, we intend to explore adopting their approach for identifying equivalence classes of program states.

Jaygarl et al [25] proposed OCAT to improve the coverage performance of the random testing tool Randoop. OCAT operates as follows: 1) It captures objects with non-isomorphic states for each class type. State isomorphism of objects is checked following Rostra [54]. 2) The objects are used as seeds for Randoop to generate more object instances, which entails randomly generating method sequences. 3) It mutates the objects in order to cover the not-yet-covered branches, which involves applying an SMT solver.

In relation to model-based testing, Mouchawrab *et al.* [43] compared testing techniques that are based on structural coverage to those based on UML state diagrams. Their results showed that there is no significant difference in terms of fault detection effectiveness, and that the two techniques are complementary. Asoudeh and Labiche [3] used a genetic algorithm to generate minimal cost test suites from finite state machines. Gao *et al.* [17] and Turner *et al.* [50] treated program objects as each having a set of states with transitions between them; the transitions are considered to be triggered when the objects' methods are invoked. They proposed test generation techniques based on such models. It is worth noting that Elbaum *et al.* [15] also modeled objects as state machines.

## VII. CONCLUSIONS AND FUTURE WORK

We presented a new state profiling approach, termed *Substate Profiling*, which is fine grained and suitable for profile-based dynamic analysis techniques. We empirically evaluated our approach using greedy test suite reduction by comparing its effectiveness to that of commonly used structural profiles. Our results showed that, in most cases, *Substate Profiling* is more or equally effective in regard to defect detection, and that it is comparable in regard to cost and to the sizes of the reduced test suites. Also, *Substate* profiles were found to be complementary to structural profiles in many cases.

In future work we will investigate ways to estimate a recommendable value of *k* by analyzing the structure of the programs under test and by mining their bug repositories. We will also apply *Substate Profiling* to other suitable techniques, such as test suite prioritization and fault localization.